\documentclass[english]{emulateapj}
\usepackage[latin9]{inputenc}
\usepackage{graphicx}

\makeatletter

\providecommand{\tabularnewline}{\\}

\makeatother

\usepackage{babel}

\begin{document}

\title{On the origin of planets at very wide orbits\\
 from the re-capture of free floating planets}

\author{Hagai B. Perets$^{1,2}$ and M.B.N. Kouwenhoven$^{3}$}

\affil{$^{1}$Harvard-Smithsonian Center for Astrophysics, 60 Garden St.,
Cambridge, MA, USA 02138\\
$^{2}$Technion - Israel Institute of Technology, Haifa, Israel\\
$^{3}$Kavli Institute for Astronomy and Astrophysics at Peking
University, Yi He Yuan Lu 5, Hai Dian District, Beijing 100871, China}

\email{hperets@physics.technion.ac.il;thijskouwenhoven@gmail.com}
\begin{abstract}
In recent years several planets have been discovered at wide orbits
($>100$ AU) around their host stars. Theoretical studies encounter
difficulties in explaining their formation and origin. Here we propose
a novel scenario for the production of planetary systems at such orbits,
through the dynamical recapture of free floating planets (FFPs) in
dispersing stellar clusters and stellar associations. This process
is a natural extension of the recently suggested scenario for the
formation of wide stellar binaries. We use N-body simulations of dispersing
clusters with $10-1000$ stars and comparable numbers of FFPs to study
this process. We find that planets are captured into wide orbits in
the typical range $\sim{\rm few}\times100-10^{6}$ AU, and
have a wide range of eccentricities (thermal distribution). Typically,
$3-6\times(f_{{\rm FFP}}/1)$ \% of all stars capture a planetary
companion with such properties 
(where $f_{{\rm FFP}}$ is the number of FFP per star in
the birth clusters). The planetary capture efficiency is comparable
to that of capture-formed stellar-binaries, and shows a similar dependence
on the cluster size and structure. It is almost
independent of the specific planetary mass; planets as well as sub-stellar
companions of any mass can be captured. The capture efficiency
decreases with increasing cluster size, and for a given cluster size the
it increases with the host/primary mass. We also find
that more than one planet can be captured around the same host through
independent consecutive captures; similarly planets can be captured
into binary systems, both in circumstellar and circumbinary orbits.
We also expect planets to be captured into pre-existing planetary
(and protoplanetary systems) as well as into orbits around black holes and massive white dwarfs, if these formed early enough before the cluster dispersal.
 In particular,
stellar black holes have a high capture efficiency ($>50$\% and $5-10\times(f_{{\rm FFP}}/1)$
\% for capture of stars and planetary companions, respectively) due
to their large mass. Finally, although rare, two FFPs or brown dwarfs
can become bound and form a FFP-binary system with no stellar host. 
\end{abstract}

\section{Introduction}

Exo-planetary systems are known to exist in a wide variety of configurations,
raising many challenges to our understanding of the complex processes
of planet formation. Some of these planetary systems were found to
have massive planets orbiting their host stars at wide orbits (hundreds
of AU; \citealp{2008ApJ...689L.153L,2011ApJ...730...39B,2011ApJ...726..113I,2011ApJ...730...42L}).
Such systems are not easily produced in current models of planet formation,
and their origin is still debated \citep[e.g. ][]{dod+09a,bos11,kra+10,ver+09}.
Microlensing observations findings show the possible existence of
a large population of apparently free floating planets (FFPs, $1.8_{-0.9}^{+1.3}$
per star; \citealp{sum+11}), i.e. they appear not to have any host
star\footnote{We do caution that the actual fraction of FFPs is unknown; it is possible that the apparently host-less planets observed have companions too faint to be resolved or too far away. We therefore scale our results throughout this paper by the FFP fraction per star. }, consistent with theoretical expectations of the production of
such runaway planets through various processes \citep[e.g.,][and references therein]{kro+03,ver+09,bea+11,par+12}. 
Here we suggest a novel scenario connecting FFPs with wide-orbits
planets, namely, the production of wide-orbit planetary systems from
the re-capture of FFPs in dispersing clusters. 

Recently, several studies have quantitatively shown that wide stellar
binaries can form following the dispersal of a stellar cluster or
stellar association (\citealp[; hereafter paper I]{2010MNRAS.404.1835K};
\citealp{2010MNRAS.404..721M,moe+11}), and a similar suggestion was
raised for the formation of Oort clouds \citep{lev+10}. Since the
majority of stars are thought to be born in star clusters \citep{2003ARA&A..41...57L},
the majority of planetary systems should also form and evolve in such
environments. Indeed, many studies explored the evolution of planets in stellar clusters (prior to the clusters dispersal), showing they could be dynamically excited by encounter with other cluster stars or even be expelled from their host star to become free floating planets (e.g. \citealp{lau+98,bon+01,smi+01,ada+06,fre+06,spu+09,mal+11,par+12} and references therein).
 Studies of the the kinematic properties of such FFPs suggest they have similar velocity dispersion as the cluster stars \citep{par+12}. The suggested scenario for the formation of wide binaries following the cluster dispersal could therefore similarly produce wide planetary systems through the same capture process, where we assume that the FFP populations is
formed prior to the cluster dispersal, which is generally consistent
with the suggested origins of such FFPs.

The capture processes can be described as follows (see also paper
I); in an evolving star cluster, stars and FFPs may
become bound to each other as the cluster expands, i.e. if the gravitational
influence of the other cluster members decreases. In order to form
a binary pair in this way, (i) they need to be sufficiently close
together, (ii) a star and a planet (or even two planets/brown dwarfs)
need to have a sufficiently small velocity difference and (iii) the
newly formed binary should not be destroyed by gravitational interaction
with the remaining cluster stars, field stars or the Galactic tide.
In the following we present the results of detailed N-body simulations
which show that wide-orbit planetary systems around both single and
binary stars can form at large numbers under realistic conditions.
In addition we note that this mechanism can produce FFP-binary systems,
brown-dwarf planetary systems such as been recently observed \citep[e.g. ][]{jay+06}
as well as planetary systems around compact objects (white dwarfs
and black holes).

The paper is organized as follows. In Section \S2 we provide a brief
overview of our technique and assumptions, which are described in
details in in Paper I. In Section \S3 we describe the results of our
simulations, and in Section \S4 we present and discuss our results. We summarize our conclusions in Section \S5.

\section{Simulations}

Generally, our models follow the same scheme as described in detail
in Paper I; in the following we briefly review the simulations. Following
Paper I, we simulate star clusters using the STARLAB package \citep{2001MNRAS.321..199P}.
We draw $N$ single stars from the \citet{2001MNRAS.322..231K}
mass function, $f_{{\rm M}}({\rm M})$, in the mass range $0.1\leq M\leq7\,{\rm M}_{\odot}$.
The lower-limit corresponds to the hydrogen-burning limit. The upper
limit is set to the mass of the most massive star which has not yet
evolved due to stellar evolution in 50 Myrs, the typical timescale
at the end of our simulations. We also run simulations with higher
cut-off ($20$ and $50$ ${\rm M}{}_{\odot}$) to probe capture by
massive stars in fast dispersing clusters.

We perform simulations with varying $N$, ranging from small stellar
systems (or sub-clumps) with $N=10$ to intermediate systems
$N=20,\,50,\,100,\,200,\:500$ and up to open cluster-sized
systems ($N=1000$). The cluster size which encloses all stars
initially is taken to be related to the number of stars through $R=0.1N^{1/3}$
pc, which corresponds to an identical initial stellar density for
all the simulated clusters. 

We study two sets of dynamical models to describe the cluster structure:
Plummer models and sub-structured (fractal) models; these are described
in detail in Paper I. Fractal (clumpy) models are characterized by
their fractal dimension $\alpha$, and are structured following the
method used by \citet{all+09}; we use two models, fractal clumpy
clusters with $\alpha=1.6$ and homogeneous systems with $\alpha=3$;
in this study we explore only initially expanding clusters with $Q=3/2$,
where $Q$ is the virial ratio, defined by $Q\equiv-E_{{\rm K}}/E_{{\rm P}}$
, where $E_{{\rm K}}$ and $E_{{\rm P}}$ are the kinetic and potential
energy of the cluster (see details in Paper I). 

All simulations are performed until the clusters are completely dissolved,
which is typically of the order of $50$ Myr (we follow the largest
clusters up to $200$ Myrs to ensure they completely dissolve), the
timescale at which the majority of low-mass star clusters are destroyed
\citep[see, e.g.][and numerous others]{1978A&A....70...57T,2003MNRAS.338..717B,2005A&A...431..905B,2005ApJ...631L.133F}.
Note that the lifetime of massive stars with $M>7$ ${\rm M}_{\odot}$
could be shorter than the timescale for the cluster dispersal, and
therefore black holes (BHs) and massive white dwarfs (WDs) may already
form in the cluster during its dispersal. Neutron stars may also form
in the cluster but are not likely to be retained in the cluster due
to their natal kicks. 

Binary systems and multiple systems (where the masses can be stellar
or planetary) are identified recursively. We impose that the systems
must be stable. The internal stability is partially guaranteed by
the fact that they survive 50 Myr in the simulations. Furthermore,
we impose the stability criterion by \citet{val+08}. In our simulations
we account for the tidal field of the galaxy at the Solar neighborhood
(following \citealp{kro+01}; we adopt a distance of $8.5$ kpc from
the Galactic center and take $M_{{\rm gal}}=5\times10^{10}$ M$_{\odot}$).
\citet{jia+10} show that binaries can survive in the Galactic tidal
field if their separations are smaller than the Jacoby radius \begin{equation}
r_{J}\sim\left(\frac{G(m_{1}+m_{2})}{4\Omega A}\right)^{1/3}\sim1.7\,\left(\frac{m_{1}+m_{2}}{{2\rm M}_{\odot}}\right)\,{\rm pc},\label{eq:Jacoby}\end{equation}
where $G$ is the gravitational constant, $\Omega$ the angular velocity
of the Galaxy at the solar circle, A is Oort\textquoteright{}s A constant,
and $m_{1}$ and $m_{2}$ are the masses of the components. The value
of 1.7 pc is valid for the canonical values for the Galactic constants
and individual masses of 1 $M_{\odot}$. We do not follow the binaries
longer than the simulation time, and therefore some of the widest
systems formed in the simulation that we consider as bound would be
slowly disrupted in the Galactic field, later on. Nevertheless, $\sim90$
\% of all the capture-formed systems we find are within their Jacoby
radius, consistent with the expectations. Note, that wide binaries
are also likely to be destroyed through encounters with other field
stars or giant molecular clouds; these are not included in our simulations.

The most important difference from previous simulations is the addition
of a population of FFPs. We simulated FFP populations with one planet
per star, $f_{{\rm FFP}}=N_{{\rm FFP}}/N_{\star}=1$,
and a more limited number of simulations with $f_{{\rm FFP}}=0.5$
and $2$. For simplicity we always make use of single mass FFPs, taken
to be Jupiter mass; $m_{{\rm {\rm FFP}}}={{\rm M}_{\rm J}}$. Since the binding
potential of a planet and a star are completely dominated by the stellar
mass, the specific choice of the planetary mass is typically not important
for their capture around stars, as long as $m_{\rm p}\ll M_{\star}$
where $m_{{\rm p}}$ and $M_{\star}$ are the typical masses
of the FFPs and the cluster stars, respectively. Nevertheless, we
also ran simulations with lower mass, 50 Earth mass planets, $m_{{\rm {\rm FFP}}}=50{\rm {\rm M}}_{\oplus}$,
to verify this. 

The parameters of the various cluster models are summarized in Table
1. Throughout the paper we refer to models by their parameters, e.g.
model F\_N20\_P0.5\_M0.16 corresponds to a fractal cluster of 20 stars
and 10 FFPs ($f_{{\rm FFP}}=0.5$) of mass $0.16$ M$_{J}$; if not
explicitly written, $f_{{\rm FFP}}=1$ and $M_{{\rm FFP}}={\rm M}_{{\rm J}}$. 

\begin{table}
\caption{Parameters of the cluster simulation runs used in our analysis }

\begin{tabular}{|c|c|}
\hline 
{\footnotesize Parameter} & {\footnotesize Value}\tabularnewline
\hline
\hline 
{\footnotesize Structure } & {\footnotesize Fractal (F; $\alpha=1.6$), }\tabularnewline
 & {\footnotesize Homogeneous (H; $\alpha=3$)}\tabularnewline
\hline 
{\footnotesize Number of stars $(N_{\star})$} & {\footnotesize 10, 20, 50,}\tabularnewline
 & {\footnotesize 100, 200, 500, 1000}\tabularnewline
\hline 
{\footnotesize Planets to stars ratio $(f_{{\rm FFP}}$)} & {\footnotesize 0.5, 1, 2}\tabularnewline
\hline 
{\footnotesize Mass Function} & {\footnotesize Kroupa MF, }\tabularnewline
 & {\footnotesize truncated at $m=7$ M$_{\odot}$}\tabularnewline
\hline 
{\footnotesize Planetary mass} & {\footnotesize ${\rm M}_{{\rm J}},$$50{\rm M}_{\oplus}$}\tabularnewline
\hline 
{\footnotesize Cluster Radius $(R)$)} & {\footnotesize $0.1N_{\star}^{1/3}$ pc}\tabularnewline
\hline 
{\footnotesize Simulation time} & {\footnotesize 50-200 Myrs}\tabularnewline
\hline
\end{tabular}
\end{table}

\section{Results}

In the following we describe the various populations of planetary
systems we obtain for the different cluster models we explore. Table
2 summarizes the results for the total fractions of multiple systems
formed in our simulations.

\textbf{\scriptsize }%
\begin{table*}
\textbf{\scriptsize \caption{\label{tab:Results}Fractions of capture-formed systems }
}{\scriptsize \par}

%\begin{spacing}{1}
\textbf{\scriptsize }\begin{tabular}{|c|c|c|c||c|c|c||c|c|c|c|}
\hline 
 & \multicolumn{3}{c||}{{\scriptsize Multiple}} & \multicolumn{3}{c||}{{\scriptsize Multiple}} & {\scriptsize Binary FFP} & \multicolumn{2}{c|}{{\scriptsize Binary star }} & {\scriptsize Triple star}\tabularnewline
 & \multicolumn{3}{c||}{{\scriptsize Stars (\%)}} & \multicolumn{3}{c||}{{\scriptsize Planets (\%)}} & {\scriptsize (\%)} & \multicolumn{2}{c|}{{\scriptsize Planets (\%)}} & {\scriptsize Planets (\% )}\tabularnewline
\hline 
{\scriptsize \# Stellar/Planetary } & {\scriptsize 2/0} & {\scriptsize 3/0} & {\scriptsize 4/0} & {\scriptsize 1/1} & {\scriptsize 1/2} & {\scriptsize 1/3} & {\scriptsize 0/2} & {\scriptsize 2/1} & {\scriptsize 2/2} & {\scriptsize 3/1}\tabularnewline
\hline 
{\scriptsize Model} &  &  &  &  &  &  &  &  &  & \tabularnewline
\hline
\hline 
\textbf{\scriptsize Homogeneous} &  &  &  &  &  &  &  &  &  & \tabularnewline
\hline 
{\scriptsize H\_N10\_P1} & {\scriptsize 10.3} & {\scriptsize 2.07} & {\scriptsize 0.12} & {\scriptsize 7.79} & {\scriptsize 0.71} & {\scriptsize 0.00} & {\scriptsize 0.01} & {\scriptsize 2.3} & {\scriptsize 0.18} & {\scriptsize 0.26}\tabularnewline
\hline 
{\scriptsize H\_N20\_P1} & {\scriptsize 8.84} & {\scriptsize 1.49} & {\scriptsize 0.10} & {\scriptsize 6.51} & {\scriptsize 0.36} & {\scriptsize 0.01} & {\scriptsize 0.00} & {\scriptsize 1.6} & {\scriptsize 0.04} & {\scriptsize 0.12}\tabularnewline
\hline 
{\scriptsize H\_N50\_P1} & {\scriptsize 6.59} & {\scriptsize 1.18} & {\scriptsize 0.10} & {\scriptsize 4.14} & {\scriptsize 0.18} & {\scriptsize 0.00} & {\scriptsize 0.00} & {\scriptsize 0.88} & {\scriptsize 0.03} & {\scriptsize 0.10}\tabularnewline
\hline 
{\scriptsize H\_N100\_P1} & {\scriptsize 5.36} & {\scriptsize 0.86} & {\scriptsize 0.09} & {\scriptsize 3.01} & {\scriptsize 0.06} & {\scriptsize 0.01} & {\scriptsize 0.00} & {\scriptsize 0.56} & {\scriptsize 0.03} & {\scriptsize 0.08}\tabularnewline
\hline 
{\scriptsize H\_N200\_P1} & {\scriptsize 4.24} & {\scriptsize 0.49} & {\scriptsize 0.03} & {\scriptsize 2.51} & {\scriptsize 0.07} & {\scriptsize 0.00} & {\scriptsize 0.00} & {\scriptsize 0.33} & {\scriptsize 0.01} & {\scriptsize 0.03}\tabularnewline
\hline 
{\scriptsize H\_N500\_P1} & {\scriptsize 2.49} & {\scriptsize 0.24} & {\scriptsize 0.02} & {\scriptsize 1.61} & {\scriptsize 0.04} & {\scriptsize 0.00} & {\scriptsize 0.00} & {\scriptsize 0.14} & {\scriptsize 0.00} & {\scriptsize 0.02}\tabularnewline
\hline 
{\scriptsize H\_N1000\_P1} & {\scriptsize 2.44} & {\scriptsize 0.16} & {\scriptsize 0.00} & {\scriptsize 1.41} & {\scriptsize 0.06} & {\scriptsize 0.00} & {\scriptsize 0.00} & {\scriptsize 0.14} & {\scriptsize 0.00} & {\scriptsize 0.01}\tabularnewline
\hline 
 &  &  &  &  &  &  &  &  &  & \tabularnewline
\hline 
{\scriptsize H\_N100\_P1\_50M$_{\oplus}$} & {\scriptsize 5.43} & {\scriptsize 0.97} & {\scriptsize 0.07} & {\scriptsize 3.16} & {\scriptsize 0.14} & {\scriptsize 0.00} & {\scriptsize 0.00} & {\scriptsize 0.64} & {\scriptsize 0.02} & {\scriptsize 0.06}\tabularnewline
\hline 
{\scriptsize H\_N100\_P0.5} & {\scriptsize 5.36} & {\scriptsize 1.15} & {\scriptsize 0.08} & {\scriptsize 1.53} & {\scriptsize 0.015} & {\scriptsize 0.00} & {\scriptsize 0.00} & {\scriptsize 0.35} & {\scriptsize 0.01} & {\scriptsize 0.04}\tabularnewline
\hline 
{\scriptsize H\_N100\_P2} & {\scriptsize 4.88} & {\scriptsize 0.70} & {\scriptsize 0.03} & {\scriptsize 5.34} & {\scriptsize 0.31} & {\scriptsize 0.00} & {\scriptsize 0.01} & {\scriptsize 0.68} & {\scriptsize 0.05} & {\scriptsize 0.11}\tabularnewline
\hline 
 &  &  &  &  &  &  &  &  &  & \tabularnewline
\hline 
\textbf{\scriptsize Fractal} &  &  &  &  &  &  &  &  &  & \tabularnewline
\hline 
{\scriptsize F\_N10\_P1} & {\scriptsize 9.97} & {\scriptsize 2.65} & {\scriptsize 0.30} & {\scriptsize 8.71} & {\scriptsize 1.02} & {\scriptsize 0.04} & {\scriptsize 0.09} & {\scriptsize 3.69} & {\scriptsize 0.23} & {\scriptsize 0.55}\tabularnewline
\hline 
{\scriptsize F\_N20\_P1} & {\scriptsize 8.69} & {\scriptsize 2.55} & {\scriptsize 0.35} & {\scriptsize 7.72} & {\scriptsize 0.68} & {\scriptsize 0.03} & {\scriptsize 0.02} & {\scriptsize 2.56} & {\scriptsize 0.16} & {\scriptsize 0.42}\tabularnewline
\hline 
{\scriptsize F\_N50\_P1} & {\scriptsize 7.42} & {\scriptsize 1.90} & {\scriptsize 0.26} & {\scriptsize 6.06} & {\scriptsize 0.51} & {\scriptsize 0.01} & {\scriptsize 0.02} & {\scriptsize 1.57} & {\scriptsize 0.15} & {\scriptsize 0.35}\tabularnewline
\hline 
{\scriptsize F\_N100\_P1} & {\scriptsize 6.79} & {\scriptsize 1.64} & {\scriptsize 0.17} & {\scriptsize 5.4} & {\scriptsize 0.37} & {\scriptsize 0.04} & {\scriptsize 0.01} & {\scriptsize 1.35} & {\scriptsize 0.09} & {\scriptsize 0.22}\tabularnewline
\hline 
{\scriptsize F\_N200\_P1} & {\scriptsize 6.32} & {\scriptsize 0.91} & {\scriptsize 0.16} & {\scriptsize 5.0} & {\scriptsize 0.27} & {\scriptsize 0.00} & {\scriptsize 0.03} & {\scriptsize 0.93} & {\scriptsize 0.04} & {\scriptsize 0.12}\tabularnewline
\hline 
{\scriptsize F\_N500\_P1} & {\scriptsize 5.24} & {\scriptsize 0.81} & {\scriptsize 0.12} & {\scriptsize 3.92} & {\scriptsize 0.22} & {\scriptsize 0.01} & {\scriptsize 0.02} & {\scriptsize 0.59} & {\scriptsize 0.04} & {\scriptsize 0.11}\tabularnewline
\hline 
{\scriptsize F\_N1000\_P1} & {\scriptsize 5.4} & {\scriptsize 0.75} & {\scriptsize 0.12} & {\scriptsize 3.94} & {\scriptsize 0.18} & {\scriptsize 0.00} & {\scriptsize 0.02} & {\scriptsize 0.55} & {\scriptsize 0.03} & {\scriptsize 0.08}\tabularnewline
\hline 
 &  &  &  &  &  &  &  &  &  & \tabularnewline
\hline 
{\scriptsize F\_N100\_P1\_50M$_{\oplus}$} & {\scriptsize 6.52} & {\scriptsize 1.52} & {\scriptsize 0.18} & {\scriptsize 5.29} & {\scriptsize 0.32} & {\scriptsize 0.02} & {\scriptsize 0.03} & {\scriptsize 1.07} & {\scriptsize 0.07} & {\scriptsize 0.27}\tabularnewline
\hline 
{\scriptsize F\_N100\_P0.5} & {\scriptsize 7.24} & {\scriptsize 1.71} & {\scriptsize 0.33} & {\scriptsize 2.78} & {\scriptsize 0.07} & {\scriptsize 0.01} & {\scriptsize 0.00} & {\scriptsize 0.72} & {\scriptsize 0.04} & {\scriptsize 0.1}\tabularnewline
\hline 
{\scriptsize F\_N100\_P2} & {\scriptsize 5.87} & {\scriptsize 1.26} & {\scriptsize 0.23} & {\scriptsize 8.51} & {\scriptsize 0.79} & {\scriptsize 0.03} & {\scriptsize 0.10} & {\scriptsize 1.45} & {\scriptsize 0.17} & {\scriptsize 0.23}\tabularnewline
\hline
\end{tabular}
%\end{spacing}

\end{table*}
{\scriptsize \par}

\subsection{Fractions of capture-formed systems}

As can be seen in Fig. 1, the fractions of capture-formed stellar-binaries
and planetary systems follow similar dependence on cluster sizes and
cluster structure (homogeneous vs. fractal). The binary fraction in
homogeneous clusters is inversely proportional to the number of stars
in the cluster, but shows a weaker dependence ($f_{{\rm cap}}\propto N_{\star}^{-0.15}$)
on the cluster size than suggested by the theoretical expectation of $f_{{\rm cap}}\sim1/N_{\star}$ from
the theory of soft binaries freeze-out \citep{moe+11}. This model predicts
a single binary to form per cluster, independent of the number of stars
in the cluster. The fractal clusters also show a decrease in capture
efficiency with increasing cluster size, with a stronger dependence, $f_{{\rm cap}}\propto N_{\star}^{-0.35}$. 

We can now integrate the observed fractions (from our simulations)
together with the observed cluster mass function $d{N}/d{N}_{\star}\propto{N}_{\star}^{-2}$\citep{2003ARA&A..41...57L}
in the range $10\leq{N}_{\star}\le1000$, and calculate an overall
expected fraction of capture-formed systems. For the homogeneous clusters
we find fractions of $f_{{\rm bin}}^{{\rm cap}}\sim0.069$ and $f_{{\rm planet}}^{{\rm cap}}\sim0.058$
for the stellar and planetary systems, respectively; similarly, for
the fractal clusters, we get $f_{{\rm bin}}^{{\rm cap}}\sim0.055$
and $f_{{\rm planet}}^{{\rm cap}}\sim0.036$. 

The observed wide binary fraction (with semi-major axis, SMA, $a>10^{4}$
AU) is $\sim0.02$ \citep{rag+10,moe+11}; recent observations focusing
on binaries at larger separations suggest that the fraction could
be even higher \citep{sha+11}, if these are accounted for. In the
simulations a large fraction ($\sim80$ \% and $\sim50$
\% for the homogeneous and fractal clusters, respectively) of the
capture-formed multiple stellar systems reside in this range. We therefore
conclude that the observed fraction of wide binaries in our simulations
is generally consistent with observations. Taken at face value, we
may therefore expect the efficiency of planetary systems formation
to be in the range of $0.036\le f_{{\rm planet}}^{{\rm cap}}\times(f_{{\rm FFP}}/1)\le0.058$,
comparable to that of capture-formed stellar systems. 

Observations of the Taurus-Auriga star-forming region show a wide
binary frequency of $\sim5-10$ \% in the range $1000-5000$ AU \citep{kra+11},
which is 2-3 times larger than found in the fractal models, and even
higher compared with the homogeneous models. Note, however, that for stars at
these separations other processes beside dynamical capture are also
likely to contribute to binary formation, and therefore direct comparison
should be taken with caution. 

\begin{figure}
\includegraphics[scale=0.45]{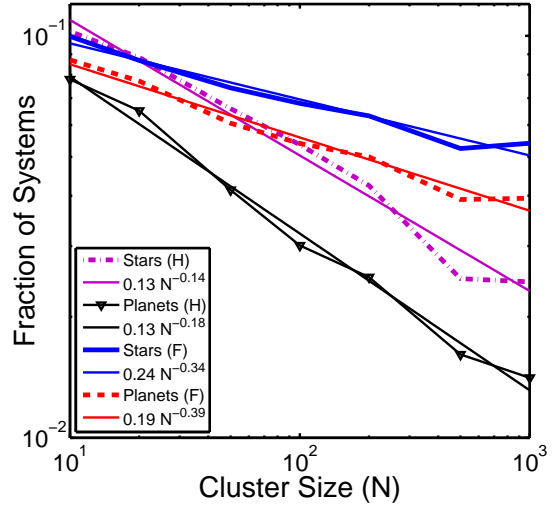}\caption{\label{fig:Fraction}Capture efficiency as a function of cluster size
(${N}_{\star}$). The fraction of capture-formed systems is a
decreasing function of the cluster size/mass, as expected from theoretical
models; however, the dependence is weaker than the $1/{N}_{\star}$dependence
predicted by models of soft binaries freeze-out \citep{moe+11}. The H and F 
models correspond to homogeneous and fractal models, respectively.}

\end{figure}

\subsection{Properties of capture-formed systems }

In the following we discuss the distributions of various properties
of the capture formed planetary and stellar systems. For simplicity
we only present results from simulations of ${N}_{\star}=100$
clusters, which generally provide a good representation of the overall
distributions we observe. 

\textbf{Mass function:} Fig. \ref{fig:host-MF} shows the fraction
of captured-formed systems as function of the primary-component mass.
As can be seen, the fraction is a monotonically increasing function
of mass (which becomes saturated at high masses), for both stellar
and planetary systems. 

\textbf{SMA and closest approach distributions:} 
The detailed distributions of SMAs and peri-center approach for
various models are shown in Fig. \ref{fig:SMA}. The orbits of captured
stellar binaries and FFPs both follow quite similar distributions.
The majority of the systems have SMAs in the typical range ${\rm few}\times10^{2}-10^{6}$ AU, with thin tails extending towards lower and larger SMAs. 
Binaries with separations larger than $\sim10^6$ AU are likely to be dissolved by the Galactic tide on timescales longer than our simulation, or through encounters with other field stars. Such systems are generally rare. 
The typical capture-formed
planetary systems found in our homogeneous cluster simulations 
follow an almost log-normal
distribution with a tail at low SMAs,
and an upper cut-off mostly due to the Galactic tide, as mentioned above. 
The fractal cluster show a log-constant distribution at the mid range
($10^{3.5}-10^{5.5}$ AU), i.e. a uniform distribution per logarithmic bin ($\ddot{{\rm O}}$pic like distribution). 
Fig. \ref{fig:SMA}b shows the peri-center distribution, 
which extends as close as a few AU from the host stars (though these extreme cases are rare). Planets and stars captured into such close approaches can have a significant dynamical interactions with pre-existing planetary/protoplanetary systems even in dynamical timescales following their capture.   
 
We note that there is a weak trend in the dependence of the
SMA and peri-center  distribution on cluster sizes. We find that 
closer binaries/planetary systems are formed in larger clusters (not shown);
 the overall distribution is generally well represented by the shown distribution in Fig. \ref{fig:SMA}.

\begin{figure}
\includegraphics[scale=0.45]{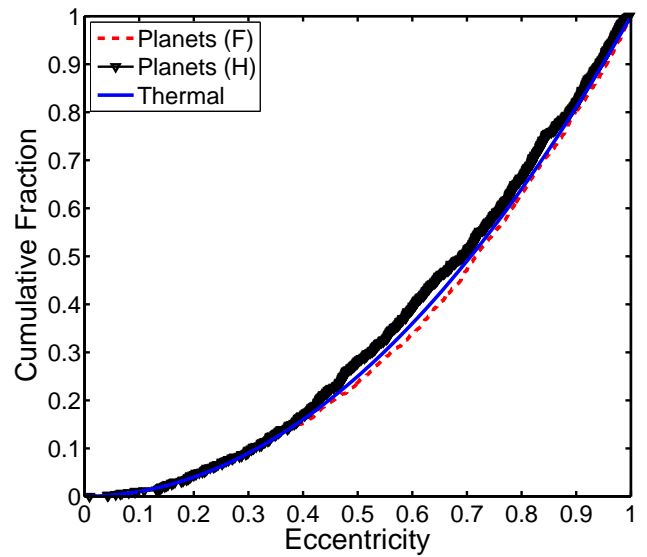}
\caption{\label{fig:ecc}The cumulative eccentricity distribution of capture-formed planetary systems. The eccentricity distribution closely follows a thermal distribution (shown in solid line).}

\end{figure}

\begin{figure}
\includegraphics[scale=0.45]{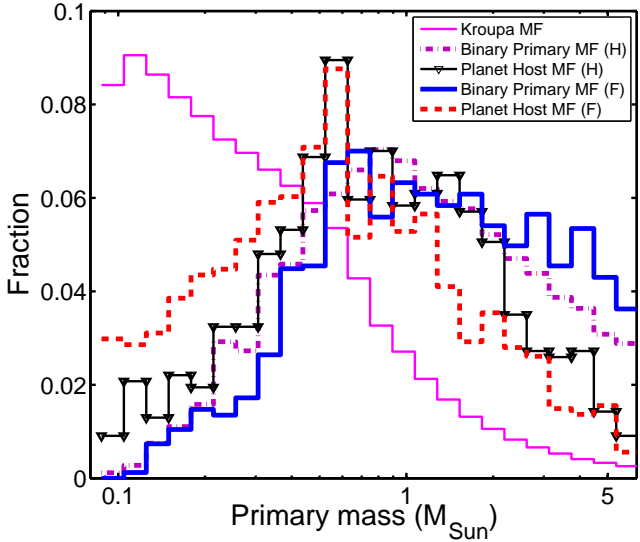}

\includegraphics[scale=0.45]{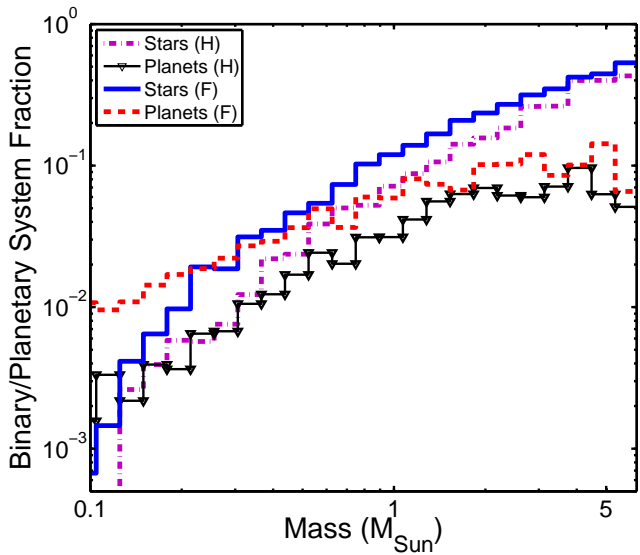}\caption{\label{fig:host-MF}The mass function of capture-formed system. Left:
The mass function of the primary star/host in capture-formed binaries/planetary
systems. Right: The fraction of capture-formed systems as a function
of the primary mass. As can be seen, the capture efficiency of both
stellar and planetary companions increases with the host mass.  The H and F 
models correspond to homogeneous and fractal models, respectively.}

\end{figure}

\begin{figure}
\includegraphics[scale=0.45]{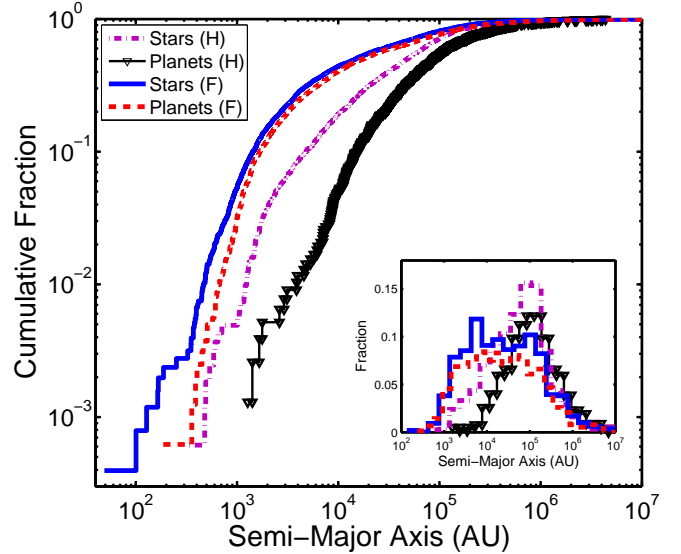}
\includegraphics[scale=0.45]{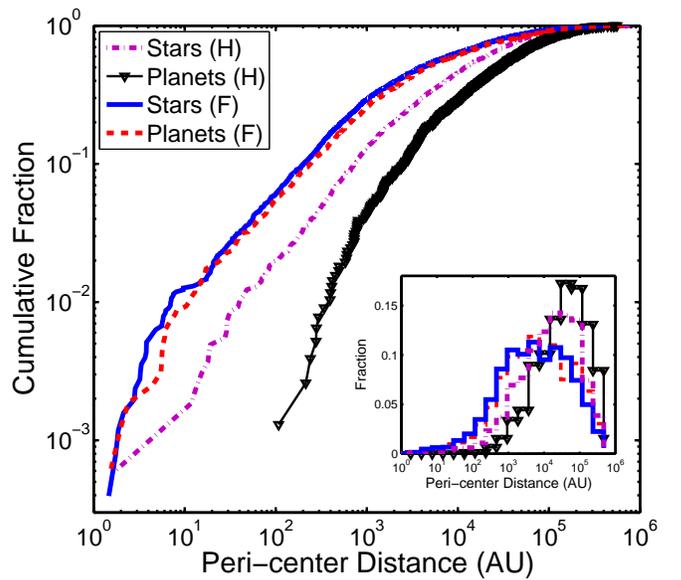}\caption{\label{fig:SMA}The semi-major axis and peri-center distribution function of capture-formed
stellar and planetary systems, in ${N}_{\star}=100$ clusters.
Top: The main figure shows the cumulative SMA distribution, and the inset shows
the differential SMA distribution. Capture formed systems from the the
fractal clusters (F) are characterized by a log-uniform distribution
(equal fraction in logarithmic bins), where as the homogeneous clusters
(H) give rise to a more log-normal distribution. 
Bottom: The same for closest approach (peri-center) distributions.  Note different axis scales.
The distribution of different cluster sizes behave similarly with a general trend for the distribution in more massive clusters to peak at smaller separations.}
\end{figure}

\textbf{Eccentricity distribution:} We find the eccentricities of
both planetary systems and stellar-binaries to follow a thermal distribution (see Fig. \ref{fig:ecc}), $d{N}/de=2e$.  

\textbf{Inclination distribution:} In higher multiplicity systems,
we can explore the relative inclinations between inner and outer binaries
of hierarchical systems. We find the inclinations distribution appears
to be isotropic (uniform in $\cos i$). 

\section{Discussion}

The cluster dispersal scenario for the formation of wide orbit planetary
systems is a natural extension for a similar model used to explain
the origin of wide binaries. Our simulation results verify that FFPs
can be captured through this process, with comparable efficiency.
The capture process depends on both the cluster properties, as well
as the masses of the specific stars. We find that the general properties
of the wide planetary systems follow the same behavior as wide binaries,
which was explored in detail in Papers I and in \citet{moe+11}. In
the following we focus on the specific implications for planetary
systems.

\subsection{Planets and brown dwarfs at wide orbits}

The most basic result of our suggested FFP-capture scenario, verified
in the detailed simulations, is the robust production of wide-orbit
planetary systems following cluster dispersal which include FFPs.
The fraction of capture-formed wide planetary systems in our simulations
is dependent on the their yet unknown original number fraction in
the birth clusters. The recent microlensing finding of FFPs, as well
as theoretical predictions, suggest that these fractions are high$^1$
$0.5-2$ FFPs per cluster star \citep{sum+11}. The FFP recapture
mechanism could therefore play a major role in the production of wide-orbit
planetary system. Such planets can regularly form around stars at
small separation, then be ejected from the system, likely through
scattering with other planets, and then be recaptured by another cluster
star following the birth-cluster dispersal. The relation between such
recaptured-planets and their new hosts could therefore be quite arbitrary.
In particular the pairing of stellar and planetary properties such
as metallicity, masses/sizes and spin-orbit relations should generally
be random, or at most reflect correlations that may exist between
the properties of any two objects in the birth cluster (e.g. related
metallicity). Even if the latter case is true, the random pairing
should be smeared out compared with the likely stronger correlation
between pairs of stars/planets originally formed in the same system. 

Planets could be captured into orbits of any eccentricity (through
high eccentricities are preferred, through the thermal eccentricity
distribution observed in our simulations), and a wide range of orbits,
with typical SMAs extending over the range ${\rm few}\times 10^2-10^{5}$ AU. At apo-center the most eccentric orbits could have twice as large separation as their SMA, and could therefore be more susceptible to disruption by the Galactic tide over time, if their SMA is comparabale to their Jacoby radius (Eq. \ref{eq:Jacoby}). Systems with the smallest separation and/or the highest eccentricities, the peri-center approach, $a(1-e)$, can become comparable to the separations of regularly formed planets at up to a few tens of AU. Objects on such orbits may strongly interact with any pre-existing planetary/protoplanetary system (see also next section). 

 In-situ formation of planets (and particularly massive gas-giants) at wide orbits as found in our simulations, where the mass density in the protoplanetary disk
become very small, is unlikely in the core-accretion model \citep{dod+09a}. 
Ejection of planets
from inner orbits to wide separations is possible. Such orbits would
initially be very eccentric and flyby encounters with other stars
would be required in order to put them into more circular orbits (similar
to the formation and evolution of the Oort cloud; \citealp{dun+87}). In comparison, the capture scenario can naturally produce both low and high eccentricity orbits. 
Planets at wide orbits, however, are not likely to survive in a cluster
environment where close encounters with other stars can destroy the
system (e.g. \citealp{par+12} and references therein). If such systems survive, or scattering occurs after the cluster dispersal it might be more difficult to currently distinguish between such a scenario and the suggested capture scenario introduced here. Nevertheless, our models provide specific predictions
for the dynamical properties of such systems, and moreover, connect
them to wide binary systems and population of free floating planets. The models therefore provide a wealth of potential signatures, which can distinguish them from different models (e.g. \citealp{ver+09}). A smoking gun signature for the capture scenario may exist
in the form of binary FFPs (see section \ref{sub:ffp-binaries}),
and the existence of planets in wide orbits around low mass stars,
which are unlikely to form through in-situ planet formation scenarios.
In the gravitational instability model for planet formation \citep{bos06}
massive gas planets may possibly form at wider orbits than in the
core-accretion scenario, but it is not clear if the process could
be efficient up to the very large separations as observed in our simulations.
In any case, such processes are not likely to form low mass rocky
planets at these separations. To summarize, in the FFP recapture model
we expect planet of any mass to be captured into very wide orbits;
systems which would be difficult to form in the currently suggested
planet formation models. We also note that the fraction of capture-formed
planetary systems, as well as the planetary masses distribution reflect
the FFP fraction and mass distribution in the birth-clusters, and
are likely to show similar orbital properties as wide stellar binaries.

\subsection{Multiple planets at wide orbits}

In principle more than a single planet can be captured into the same
planetary system, through independent captures. The fraction of such
multiple captured-planet systems is therefore expected to be proportional
to the probability for multiple independent captures, i.e. $f_{2{\rm p}}=f_{{\rm p}}^{2}$
for capture of two planets. Table 2 show the fraction of planetary
systems with $n=1,2$ and $3$, which is generally consistent with
the expectation. Capture of high ($>2)$ multiplicity stellar systems
is expected to follow the same behavior, however, capture of an additional
star also contributes to the total mass of the system, and therefore
the likelihood of a consecutive capture is somewhat increased. This
is verified in the higher fraction of high multiplicity stellar systems
in comparison to the planetary systems (when normalized to the fraction
of binary/single planet systems). 

Though this scenario can explain even multiple planet systems at wide
orbits, it is unlikely to form a large number of such systems, given
the rapidly decreasing probability for such occurrence. It is also
unlikely to form planetary systems in mean motion resonance, which
require a continuous migration of planetary orbits through phase space;
these are likely to form through migration in disks. 

Naturally, stars hosting primordial in-situ formed planetary systems
or protoplanetary disk can capture additional planets following the
cluster dispersal. Given the random orientation of the capture processes,
these additional 'guest' captured planets are likely to be misaligned
with any existing protoplanetary disk or pre-existing planetary system.
In some cases, the capture of an additional object into a wide orbit
may lead to the perturbation or even the destabilization of an existing
planetary system. In particular, an object captured at high inclination
may induce long term secular evolution of the inner pre-existing system
(e.g. through Kozai-Lidov evolution; \citealp{koz62,lid62}), and
potentially even destabilize them (though the latter possibility is
far more likely in the case of a stellar capture).

\subsection{Planets in binary systems: circumstellar and circumbinary orbits}

Primordial binaries existing before the cluster dispersal could themselves
capture additional stellar or planetary objects, forming high multiplicity
stellar systems of binary planets. This mechanism is not much different
than the capture of planet around single stars (aside from the potentially
much higher mass of the system, in the case of binary stars, which
induces a higher capture probability, see Fig. \ref{fig:host-MF}).
Nevertheless, three body systems could become unstable, and therefore
the capture of additional objects by a binary may limit the range
of orbital phase-space available, at least in those cases where the
initial binary is wide. The current simulations presented here did
not include primordial binaries, which would be a natural extension
of our models. We expect such systems to generally have a higher capture
efficiency due to their higher mass.

Binary planets could also form through a double capture; either a
planet capture followed by a stellar capture or a stellar capture,
forming a binary, followed by a planetary capture. The planetary capture
in the latter case, follows a similar evolution as planetary capture
around primordial binaries; however, the inner stellar binary would
be typically much wider than primordial binaries in the birth cluster.
The former process where a planet is captured first, may lead to a
circumstellar binary planet, as the stellar binary could be captured
at a much wider orbit, not perturbing the inner planetary orbit. In
this case, \emph{circumbinary} planetary systems is less likely to
form, as a close single-binary encounter of a star within the the
planetary orbit (i.e. where the binary is the star-planet system),
would more likely lead to its ejection process (typically the lowest
mass objects are ejected in such encounters). Table 2 shows the fraction
of binary planets (combining both circumstellar and circumbinary orbits)
formed in our simulations.

\subsection{Free floating planet-binaries}

\label{sub:ffp-binaries}

Larger stellar masses allow for binding of companions with higher
relative velocities, as reflected in the higher capture rate by more
massive stars. Nevertheless, planets could capture companion planets
without any host stars. These free floating double planets or FFP-binaries,
are generally rare (see Table 2). The formation of such systems (or
similarly planet capture around brown dwarfs) through typical planet
formation scenarios rather than through dynamical capture processes
is unlikely (current models require the existence of a host stars
around which planet forms in a massive enough protoplanetary disk;
see also discussion on planet and star formation at wide orbits in
\citealt{bej+08,2009ApJ...703.1511K}). Therefore, the finding of
even a single such system could serve as a potential smoking gun signature
of the planetary capture scenario discussed here. Given the increased capture probability with system mass, and the better detectability of more massive planets, the most likely FFP-binaries to be observed would lie in the higher end of the planetary mass regime. Most interestingly,
\citet{jay+06} discovered such a systems of two planetary mass companions of a few$\times10\,{\rm M_J}$ at wide, 240 AU separation, followed by the discovery of similar even
wider systems by other groups \citep{bej+08,2011ApJ...730...39B}.
Our suggested scenario can provide a natural explanation for these
otherwise puzzling systems.

\subsection{Wide orbit planets around compact objects}

The timescales for cluster dispersal could be relatively short, and
much shorter than the main-sequence lifetimes of most stars. Nevertheless,
the lifetime of stars more massive that $\sim5$ ${\rm M}_{\odot}$
could be comparable to the dispersal time scales of the host cluster.
These stars could therefore become compact objects before the cluster
disperses, and can later capture planets like any other cluster star.
Note, however that this is likely to be true only for black holes
and white dwarfs; the natal kicks of neutron stars are likely to ejects
them from their host cluster upon their formation. The mass loss of
stars prior to the formation of the compact object can eject pre-existing
planets at wide orbits (e.g. Veras et al. 2012, submitted); especially
in the case of the prompt mass loss in a supernova. The capture channel
suggest that even such objects could host wide-orbit planetary systems.
Interestingly, the dispersal timescale of a cluster defines a lower
limit on the mass of WD progenitors that could acquire planetary/binary
companions through capture. Lower mass WDs with such wide orbits companion
would have had to capture them while still on the MS, and would have
a higher chance of losing these companions as they evolve and lose
mass. The age cut-off may therefore be reflected by a lower fraction
of low mass WDs with wide orbit stellar or planetary companions. Masses
of stellar black holes could typically be in the range of $5-15$
M$_{\odot}$; given the higher probability for planetary and stellar
capture for massive stars (see Fig. \ref{fig:host-MF}), we expect
more than half of all stellar black holes to have a wide orbit captured
stellar companion, and more than $5-10\times(f_{{\rm FFP}}/1)$ \%
of them to host a planetary companion (our additional simulations
of clusters containing stars as massive as $20$ ${\rm M}_{\odot}$,
not shown here, suggest that the capture probability for stellar companions
could approach unity for $m_{\star}>10$ M$_{\odot}$). Massive white
dwarfs ($m_{{\rm WD}}\sim1$ M$_{\odot}$) are expected to capture $\sim0.2$ less companions compared with stellar black holes.

\section{Summary}

In this paper we presented a novel channel for the formation of planetary
systems at very wide orbits (typically $\sim{\rm few}\times10^{2}-10^{5}$ AU), through the
dynamical recapture of free floating planets during a cluster dispersal.
This mechanism, originally suggested as a route for the formation
of wide orbit binaries, was extended to the planetary regime. We simulated
dispersing clusters with population of FFPs, and followed the formation
of planetary systems in the simulations. We find that the efficiency
of capture-formation of planetary systems ranges between $1-9$ \%
depending on cluster size and structure. We predict and overall fraction
of $3-6\times(f_{{\rm FFP}}/1)$ \% of all stars (accounting for the birth cluster size distribution) to host a wide orbit
planetary companion at separations $>100$ AU, 
comparable to the expected fraction of wide stellar 
binaries. The orbital properties of capture-formed planetary systems
resemble those of wide binaries. The fraction of capture-formed planetary
systems (as well as capture-formed binaries) increases with the host
mass.  Future searches for wide-orbit planetary companions should therefore favor more massive stars. Although captures are biased towards higher
mass hosts, planets can be captured around hosts of any mass (with stars like our Sun having a $3-5$ \% probability to have captured a wide orbit planet). 
In fact, we find that, although rare, planets could even be captured around
other planets (or brown dwarfs) to form free floating planet-binaries.
In addition, planets can be captured around compact objects, if these
formed prior to the cluster dissolution; we find the stellar black
holes may have a high probability of capturing stellar and planetary
companions in wide orbits. Finally, we note that the properties of
captured planets should not correlate with their host star; single
and multiple planets could be captured into arbitrary orbits, generally
non-coplanar with each other (or with the stellar host spin, or pre-existing
planetary system). Nevertheless, some secondary planet-host star metallicity
correlation may still exist due to the common origin of the stars
from same birth cluster.

\acknowledgements{HBP is a CfA and BIKURA (FIRST) fellow. M.B.N.K. was supported by
the Peter and Patricia Gruber Foundation through the IAU-PPGF fellowship,
by the Peking University One Hundred Talent Fund (985), and by the
National Natural Science Foundation of China (grants 11010237, 11050110414, 11173004). The authors would like to thank Michael Ireland, Nathan
Kaib, Sally Dodson-Robinson and the anonymous referee for helpful comments 
that helped improve this manuscript. }

\bibliographystyle{apj}
%\bibliography{planet-formation,wide-binaries}

\begin{thebibliography}{44}
\expandafter\ifx\csname natexlab\endcsname\relax\def\natexlab#1{#1}\fi

\bibitem[{{Adams} {et~al.}(2006){Adams}, {Proszkow}, {Fatuzzo}, \&
  {Myers}}]{ada+06}
{Adams}, F.~C., {Proszkow}, E.~M., {Fatuzzo}, M., \& {Myers}, P.~C. 2006, \apj,
  641, 504

\bibitem[{{Allison} {et~al.}(2009){Allison}, {Goodwin}, {Parker}, {de Grijs},
  {Portegies Zwart}, \& {Kouwenhoven}}]{all+09}
{Allison}, R.~J., {Goodwin}, S.~P., {Parker}, R.~J., {de Grijs}, R., {Portegies
  Zwart}, S.~F., \& {Kouwenhoven}, M.~B.~N. 2009, \apjl, 700, L99

\bibitem[{{Bastian} {et~al.}(2005){Bastian}, {Gieles}, {Lamers}, {Scheepmaker},
  \& {de Grijs}}]{2005A&A...431..905B}
{Bastian}, N., {Gieles}, M., {Lamers}, H.~J.~G.~L.~M., {Scheepmaker}, R.~A., \&
  {de Grijs}, R. 2005, \aap, 431, 905

\bibitem[{{Beauge} \& {Nesvorny}(2011)}]{bea+11}
{Beauge}, C. \& {Nesvorny}, D. 2011, ArXiv:1110.4392

\bibitem[{{B{\'e}jar} {et~al.}(2008){B{\'e}jar}, {Zapatero Osorio},
  {P{\'e}rez-Garrido}, {{\'A}lvarez}, {Mart{\'{\i}}n}, {Rebolo},
  {Vill{\'o}-P{\'e}rez}, \& {D{\'{\i}}az-S{\'a}nchez}}]{bej+08}
{B{\'e}jar}, V.~J.~S., {Zapatero Osorio}, M.~R., {P{\'e}rez-Garrido}, A.,
  {{\'A}lvarez}, C., {Mart{\'{\i}}n}, E.~L., {Rebolo}, R.,
  {Vill{\'o}-P{\'e}rez}, I., \& {D{\'{\i}}az-S{\'a}nchez}, A. 2008, \apjl, 673,
  L185

\bibitem[{{Biller} {et~al.}(2011){Biller}, {Allers}, {Liu}, {Close}, \&
  {Dupuy}}]{2011ApJ...730...39B}
{Biller}, B., {Allers}, K., {Liu}, M., {Close}, L.~M., \& {Dupuy}, T. 2011,
  \apj, 730, 39

\bibitem[{{Bonnell} {et~al.}(2001){Bonnell}, {Smith}, {Davies}, \&
  {Horne}}]{bon+01}
{Bonnell}, I.~A., {Smith}, K.~W., {Davies}, M.~B., \& {Horne}, K. 2001, \mnras,
  322, 859

\bibitem[{{Boss}(2006)}]{bos06}
{Boss}, A.~P. 2006, ApJL, 637, L137

\bibitem[{{Boss}(2011)}]{bos11}
---. 2011, \apj, 731, 74

\bibitem[{{Boutloukos} \& {Lamers}(2003)}]{2003MNRAS.338..717B}
{Boutloukos}, S.~G. \& {Lamers}, H.~J.~G.~L.~M. 2003, \mnras, 338, 717

\bibitem[{{Dodson-Robinson} {et~al.}(2009){Dodson-Robinson}, {Veras}, {Ford},
  \& {Beichman}}]{dod+09a}
{Dodson-Robinson}, S.~E., {Veras}, D., {Ford}, E.~B., \& {Beichman}, C.~A.
  2009, \apj, 707, 79

\bibitem[{{Duncan} {et~al.}(1987){Duncan}, {Quinn}, \& {Tremaine}}]{dun+87}
{Duncan}, M., {Quinn}, T., \& {Tremaine}, S. 1987, \aj, 94, 1330

\bibitem[{{Fall} {et~al.}(2005){Fall}, {Chandar}, \&
  {Whitmore}}]{2005ApJ...631L.133F}
{Fall}, S.~M., {Chandar}, R., \& {Whitmore}, B.~C. 2005, \apjl, 631, L133

\bibitem[{{Fregeau} {et~al.}(2006){Fregeau}, {Chatterjee}, \& {Rasio}}]{fre+06}
{Fregeau}, J.~M., {Chatterjee}, S., \& {Rasio}, F.~A. 2006, \apj, 640, 1086

\bibitem[{{Ireland} {et~al.}(2011){Ireland}, {Kraus}, {Martinache}, {Law}, \&
  {Hillenbrand}}]{2011ApJ...726..113I}
{Ireland}, M.~J., {Kraus}, A., {Martinache}, F., {Law}, N., \& {Hillenbrand},
  L.~A. 2011, \apj, 726, 113

\bibitem[{{Jayawardhana} \& {Ivanov}(2006)}]{jay+06}
{Jayawardhana}, R. \& {Ivanov}, V.~D. 2006, Science, 313, 1279

\bibitem[{{Jiang} \& {Tremaine}(2010)}]{jia+10}
{Jiang}, Y.-F. \& {Tremaine}, S. 2010, \mnras, 401, 977

\bibitem[{{Kouwenhoven} {et~al.}(2010){Kouwenhoven}, {Goodwin}, {Parker},
  {Davies}, {Malmberg}, \& {Kroupa}}]{2010MNRAS.404.1835K}
{Kouwenhoven}, M.~B.~N., {Goodwin}, S.~P., {Parker}, R.~J., {Davies}, M.~B.,
  {Malmberg}, D., \& {Kroupa}, P. 2010, \mnras, 404, 1835

\bibitem[{{Kozai}(1962)}]{koz62}
{Kozai}, Y. 1962, AJ, 67, 591

\bibitem[{{Kratter} {et~al.}(2010){Kratter}, {Murray-Clay}, \&
  {Youdin}}]{kra+10}
{Kratter}, K.~M., {Murray-Clay}, R.~A., \& {Youdin}, A.~N. 2010, \apj, 710,
  1375

\bibitem[{{Kraus} \& {Hillenbrand}(2009)}]{2009ApJ...703.1511K}
{Kraus}, A.~L. \& {Hillenbrand}, L.~A. 2009, \apj, 703, 1511

\bibitem[{{Kraus} {et~al.}(2011){Kraus}, {Ireland}, {Martinache}, \&
  {Hillenbrand}}]{kra+11}
{Kraus}, A.~L., {Ireland}, M.~J., {Martinache}, F., \& {Hillenbrand}, L.~A.
  2011, \apj, 731, 8

\bibitem[{{Kroupa}(2001)}]{2001MNRAS.322..231K}
{Kroupa}, P. 2001, \mnras, 322, 231

\bibitem[{{Kroupa} {et~al.}(2001){Kroupa}, {Aarseth}, \& {Hurley}}]{kro+01}
{Kroupa}, P., {Aarseth}, S., \& {Hurley}, J. 2001, \mnras, 321, 699

\bibitem[{{Kroupa} \& {Bouvier}(2003)}]{kro+03}
{Kroupa}, P. \& {Bouvier}, J. 2003, \mnras, 346, 369

\bibitem[{{Lada} \& {Lada}(2003)}]{2003ARA&A..41...57L}
{Lada}, C.~J. \& {Lada}, E.~A. 2003, \araa, 41, 57

\bibitem[{{Lafreni{\`e}re} {et~al.}(2011){Lafreni{\`e}re}, {Jayawardhana},
  {Janson}, {Helling}, {Witte}, \& {Hauschildt}}]{2011ApJ...730...42L}
{Lafreni{\`e}re}, D., {Jayawardhana}, R., {Janson}, M., {Helling}, C., {Witte},
  S., \& {Hauschildt}, P. 2011, \apj, 730, 42

\bibitem[{{Lafreni{\`e}re} {et~al.}(2008){Lafreni{\`e}re}, {Jayawardhana}, \&
  {van Kerkwijk}}]{2008ApJ...689L.153L}
{Lafreni{\`e}re}, D., {Jayawardhana}, R., \& {van Kerkwijk}, M.~H. 2008, \apjl,
  689, L153

\bibitem[{{Laughlin} \& {Adams}(1998)}]{lau+98}
{Laughlin}, G. \& {Adams}, F.~C. 1998, \apjl, 508, L171

\bibitem[{{Levison} {et~al.}(2010){Levison}, {Duncan}, {Brasser}, \&
  {Kaufmann}}]{lev+10}
{Levison}, H.~F., {Duncan}, M.~J., {Brasser}, R., \& {Kaufmann}, D.~E. 2010,
  Science, 329, 187

\bibitem[{{Lidov}(1962)}]{lid62}
{Lidov}, M.~L. 1962, Planetary and Space Science, 9, 719

\bibitem[{{Malmberg} {et~al.}(2011){Malmberg}, {Davies}, \& {Heggie}}]{mal+11}
{Malmberg}, D., {Davies}, M.~B., \& {Heggie}, D.~C. 2011, \mnras, 411, 859

\bibitem[{{Moeckel} \& {Bate}(2010)}]{2010MNRAS.404..721M}
{Moeckel}, N. \& {Bate}, M.~R. 2010, \mnras, 404, 721

\bibitem[{{Moeckel} \& {Clarke}(2011)}]{moe+11}
{Moeckel}, N. \& {Clarke}, C.~J. 2011, \mnras, 415, 1179

\bibitem[{{Parker} \& {Quanz}(2012)}]{par+12}
{Parker}, R.~J. \& {Quanz}, S.~P. 2012, \mnras, 419, 2448

\bibitem[{{Portegies Zwart} {et~al.}(2001){Portegies Zwart}, {McMillan}, {Hut},
  \& {Makino}}]{2001MNRAS.321..199P}
{Portegies Zwart}, S.~F., {McMillan}, S.~L.~W., {Hut}, P., \& {Makino}, J.
  2001, \mnras, 321, 199

\bibitem[{{Raghavan} {et~al.}(2010){Raghavan}, {McAlister}, {Henry}, {Latham},
  {Marcy}, {Mason}, {Gies}, {White}, \& {ten Brummelaar}}]{rag+10}
{Raghavan}, D., {McAlister}, H.~A., {Henry}, T.~J., {Latham}, D.~W., {Marcy},
  G.~W., {Mason}, B.~D., {Gies}, D.~R., {White}, R.~J., \& {ten Brummelaar},
  T.~A. 2010, \apjs, 190, 1

\bibitem[{{Shaya} \& {Olling}(2011)}]{sha+11}
{Shaya}, E.~J. \& {Olling}, R.~P. 2011, \apjs, 192, 2

\bibitem[{{Smith} \& {Bonnell}(2001)}]{smi+01}
{Smith}, K.~W. \& {Bonnell}, I.~A. 2001, \mnras, 322, L1

\bibitem[{{Spurzem} {et~al.}(2009){Spurzem}, {Giersz}, {Heggie}, \&
  {Lin}}]{spu+09}
{Spurzem}, R., {Giersz}, M., {Heggie}, D.~C., \& {Lin}, D.~N.~C. 2009, \apj,
  697, 458

\bibitem[{{Sumi} {et~al.}(2011)}]{sum+11}
{Sumi}, T. {et~al.} 2011, \nat, 473, 349

\bibitem[{{Tutukov}(1978)}]{1978A&A....70...57T}
{Tutukov}, A.~V. 1978, \aap, 70, 57

\bibitem[{{Valtonen} {et~al.}(2008){Valtonen}, {Myll{\"a}ri}, {Orlov}, \&
  {Rubinov}}]{val+08}
{Valtonen}, M., {Myll{\"a}ri}, A., {Orlov}, V., \& {Rubinov}, A. 2008, in IAU
  Symposium, Vol. 246, IAU Symposium, ed. {E.~Vesperini, M.~Giersz, \&
  A.~Sills}, 209--217

\bibitem[{{Veras} {et~al.}(2009){Veras}, {Crepp}, \& {Ford}}]{ver+09}
{Veras}, D., {Crepp}, J.~R., \& {Ford}, E.~B. 2009, \apj, 696, 1600

\end{thebibliography}

\end{document}